\documentclass[aps,prb,twocolumn,epsfig]{revtex4}
\usepackage{graphics,graphicx,epsfig,color,hyperref,latexsym,float}

\begin{document}

\author{Rajarshi Tiwari and Pinaki Majumdar}


\affiliation{
  Harish-Chandra  Research Institute,\\
  Chhatnag Road, Jhusi, Allahabad 211019, India
}

\title{Mott Transition and Glassiness in the Face Centered Cubic Lattice}

\date{12 Feb 2013}

\begin{abstract}
We study the half-filled Hubbard model on the geometrically frustrated
face centered cubic (FCC) lattice, using an auxiliary field based real 
space technique. The low temperature state is a paramagnetic metal at 
weak interaction, an antiferromagnetic insulator (AFI) with flux like 
order at intermediate interaction, and an AFI with `C type' order at 
very strong interaction. Remarkably, there is a narrow window between 
the paramagnetic metal and the AFI where the system exhibits spin glass 
behaviour arising from the presence of disordered but `frozen' local 
moments. The spin glass state is metallic at weaker interaction but shows
crossover to pseudogap behaviour and an insulating 
resistivity with growing interaction. We compare our results to 
available experiments on FCC and pyrochlore based materials and suggest
that several of these features are typical of three dimensional correlated
systems with geometric frustration.
\end{abstract}

\keywords{face centered cubic lattice, Mott transition, spin glass}
\maketitle

The presence of geometric frustration in an
interacting electron system disfavours long range magnetic 
order and promotes a complex electronic state with
short range correlations\cite{frust-rev1,frust-rev2}.
Such effects have been intensely explored 
in the context of the quasi two dimensional (2D) 
organic salts\cite{mckz,kanoda}
where, in some compounds, the triangular 
lattice structure gives rise to a spin liquid 
rather than conventional Neel order.

Unlike in 2D, there is no organised body of work
probing the interplay of geometric frustration and
Mott physics in three dimensions (3D).
There are intriguing experimental 
results on disparate 
systems 
\cite{148-jacs,148-Nb-flux,148-res-elm,148-opt-phu,c60-phd,c60-rev,c60-nat,c60-nmr,dp-CaRe,dp-LiRuYRu,dp-LiReYRe,dp-YMo,dp-InRe,pyr-Mo-sf1,pyr-Mo-sf2,pyr-Mo,pyr-Ir-msl,pyr-Ir-jpsj,pyr-Ir-anHall,pyr-Ir-cdyn,pyr-Ir}, 
whose common features do not
seem to have been noticed.
The 3D frustrated Mott systems are realised on face centered
cubic (FCC) 
and pyrochlore lattices. They both involve corner sharing
tetrahedra, disfavouring simple Neel order in the
insulating phase. 
The FCC examples include the cluster compounds 
\cite{148-jacs,148-Nb-flux,148-res-elm,148-opt-phu} like GaTa$_4$Se$_8$, GaNb$_4$Se$_8$, {\it etc},
some alkali fulleride's of the form 
\cite{c60-phd,c60-rev,c60-nat,c60-nmr} A$_3$C$_{60}$, 
and the `B site ordered' double perovskites 
\cite{dp-CaRe,dp-LiRuYRu,dp-LiReYRe,dp-YMo,dp-InRe},
{\it  e.g.}, Sr$_2$InReO$_6$. The pyrochlore examples include the 
molybdates \cite{pyr-Mo-sf1,pyr-Mo-sf2,pyr-Mo}
R$_2$Mo$_2$O$_7$ and iridates 
\cite{pyr-Ir-msl,pyr-Ir-jpsj,pyr-Ir-anHall,pyr-Ir-cdyn,pyr-Ir}
Ln$_2$Ir$_2$O$_7$. Most of these materials, at ambient `pressure'
\cite{chem-press}, are insulators close to a Mott transition.

While there is great variation among these materials, the following
features seem to be shared:
(a).~In the Mott phase they usually exhibit no long range order
down to the lowest temperature, sometimes with a hint of
`spin freezing'\cite{148-jacs,dp-InRe,dp-LiRuYRu,dp-LiReYRe,pyr-Mo-sf1,pyr-Mo-sf2}.
(b).~On pressure driven metallisation, the 
resistivity is {\it very large but finite}
\cite{148-res-elm,pyr-Mo,pyr-Ir-cdyn,pyr-Ir} at low temperature 
over a wide
pressure window, and exhibits a negative temperature
derivative, before eventual `normal' behaviour. (c)~Optical
conductivity\cite{148-opt-phu,pyr-Ir-cdyn}, where available across the pressure driven
transition, indicates large transfer of spectral weight.
(d)~The Hall conductance\cite{pyr-Ir-anHall} has a spontaneous anomalous contribution
indicating significant non coplanar character in the magnetic
background.  In addition,
(e).~some of these systems exhibit superconductivity
at low temperature\cite{148-res-elm,c60-phd,c60-rev,c60-nat,c60-nmr}.

\begin{figure}[b]
\centerline{
\includegraphics[width=6.8cm,height=5cm]{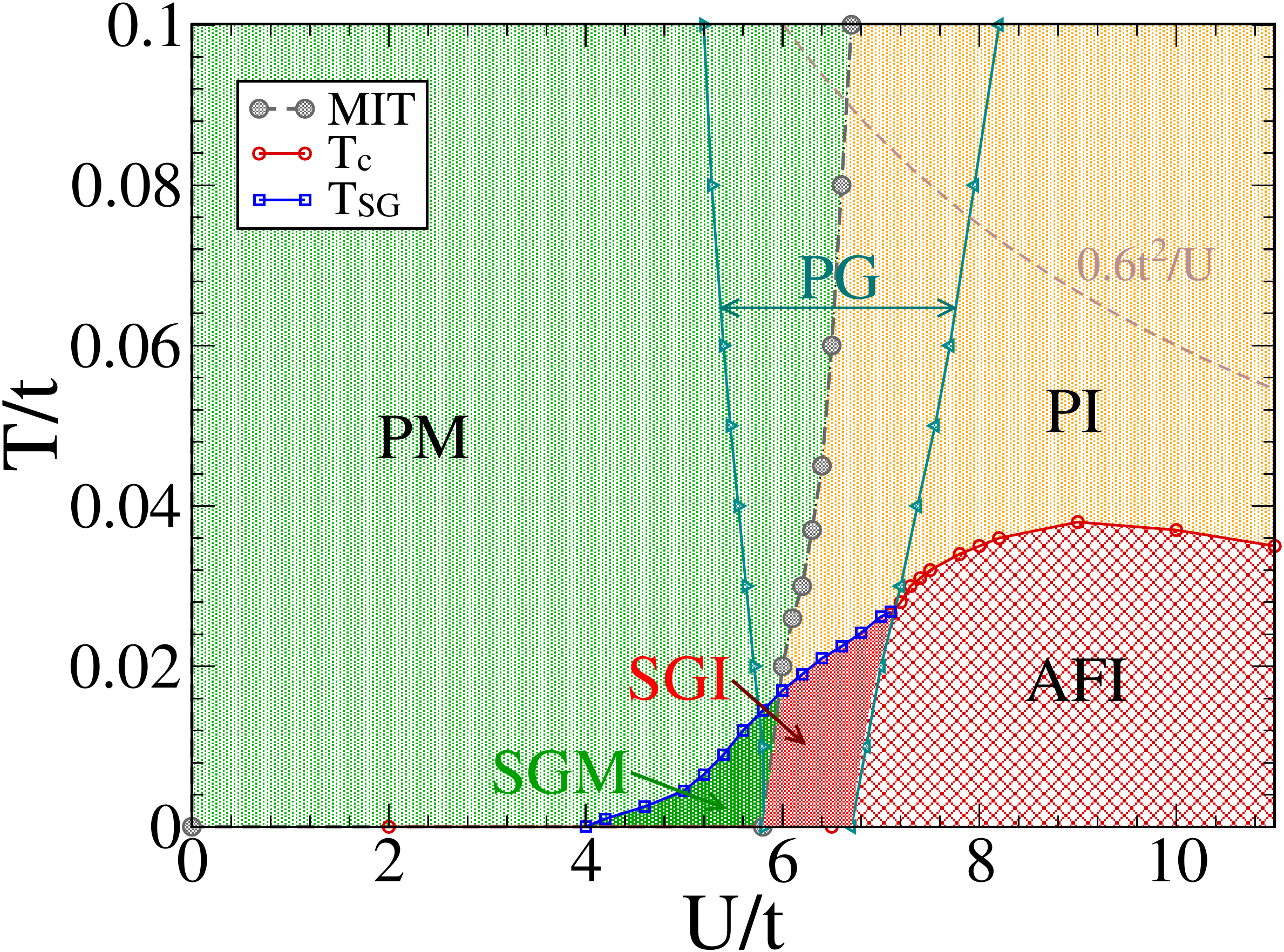}
}
\caption{\label{phd-fcc}
The $U-T$ phase diagram of the Hubbard model on the FCC lattice
at half-filling. The ground state can be a paramagnetic metal (PM), with
no local moments, a spin glass metal (SGM) or spin glass insulator
(SGI) with disordered local moments, and an antiferromagnetic
insulator (AFI). The AFI has `flux' like order upto
$U/t \approx 30$, beyond which it has `C type' order.
The classification into `metal' and `insulator' is based on the
temperature dependence of the resistivity, $\rho(T)$. $d\rho/dT >0$
indicates a metal, $d\rho/dT <0$ an insulator.
At finite temperature the system also has a paramagnetic (Mott)
insulating (PI) phase.
The magnetic transition temperature,  $T_c$, and the spin glass
freezing temperature, $T_{SG}$ (see text), are indicated. We show
the extrapolation of the $T_c \sim 0.6t^2/U$ asymptote, that describes
the $U/t \gg 1$ transition, to highlight the large 
deviation from the short range Heisenberg result.
The  PG region involves a pseudogap in the density of states.
}
\end{figure}

The complexity of the three dimensional structures has
prevented a detailed exploration of the possible
magnetic phases and the Mott transition in these
systems.
While each material involves its specific electronic
model, 
in this paper we focus on the single band Hubbard model at
half-filling on the FCC lattice. 
We use an approach that captures the unrestricted Hartee-Fock
(UHF) state at zero temperature, but retains the crucial 
thermal fluctuations of the interaction induced 
`local moments', and their impact on electronic properties.
Using a combination of Monte Carlo (MC)  and variational schemes
we establish the following:
($i$)~In the ground state, 
increasing interaction leads, successively, 
to transition from a paramagnetic metal 
(PM) to a spin glass metal (SGM), a spin glass `insulator'
(SGI), and then an antiferromagnetic
insulator (AFI). The AFI has 
flux like order at weaker coupling, and `C type' order
(see Supplement) in the very strong
coupling Heisenberg limit.   
($ii$)~The $T_c$ scales in the ordered phase, as
well as the notional
glass transition temperature $T_{SG}$, 
are a tiny fraction, $\sim$ a few percent, of the
hopping scale due to the frustration. ($iii$)~While the AFI 
phase has a clear gap and divergent resistivity 
as temperature $ T \rightarrow 0$, 
the spin glass insulator has a pseudogap (PG) in the single
particle density of states (DOS), 
non Drude optical response, and
{\it large but finite resistivity} at
$T=0$ with $d\rho/dT < 0$. ($iv$)~The transport 
trends match remarkably with experiments on 
FCC Mott systems, and allow us to make predictions
about their magnetic and spectral properties.

There is surprisingly little theoretical 
work on the magnetic phases
or the Mott transition in the FCC lattice. There is a
very early calculation exploring a restricted set of
mean field states\cite{mf-fcc} but, in contrast to
two dimensions, there does not seem to be any
cellular dynamical mean field theory (C-DMFT) result
handling the combination of correlation and frustration.
We use an approach, suggested long back by 
Hubbard himself \cite{hubb},
for the model:
\begin{equation}
H=\sum_{\langle ij\rangle\sigma}t_{ij} 
c^{\dagger}_{i\sigma}c_{j\sigma} -\mu\sum_i n_i
  + U\sum_{i}n_{i\uparrow}n_{i\downarrow}
\end{equation}
The $t_{ij}=-t$ for nearest neighbour hopping on the FCC
lattice. 
We will set $t=1$ as the reference energy scale.
$\mu$ controls the electron density, which we maintain 
at $n=1$. $U >0$ is the Hubbard repulsion.

We use a Hubbard-Stratonovich (HS)
transformation \cite{hubb-strat}
that introduces a vector field ${\bf m}_i(\tau)$ and a scalar
field $\phi_i(\tau)$ at each site to decouple the interaction.
This decomposition \cite{hubb,schulz} retains the rotation
invariance of the Hubbard model and reproduces UHF theory
at saddle point.
We treat the ${\bf m}_i$ and $\phi_i$
as classical fields, {\it i.e}, neglect their time dependence,
but completely retain the thermal fluctuations in
${\bf m}_i$.  $\phi_i$ is treated at the saddle point level, 
{\it i.e}, $\phi_i \rightarrow \langle \phi_i \rangle = (U/2)
\langle \langle n_i \rangle \rangle = U/2$ at half-filling, since
charge fluctuations would be penalised at temperatures $T \ll U$.
Retaining the spatial fluctuations of ${\bf m}_i$
allows us to estimate $T_c$ scales, and access the crucial thermal
effects on transport.
We will discuss the limitations
of the method later in the paper.

With this approximation the half-filled Hubbard problem is 
mapped on to electrons coupled to field, ${\bf m}_i$. 
\begin{equation}
H_{eff} 
=\sum_{ij,\sigma}t_{ij} c^{\dagger}_{i\sigma}c_{j\sigma}
- {\tilde \mu} N 
- \frac{U}{2}\sum_{i}{\bf m}_{i}\cdot\vec{\sigma}_{i}
+ \frac{U}{4}\sum_{i}{\bf m}_{i}^{2}
\end{equation}
where $\tilde mu = \mu - U/2$.
We can write this as $H_{eff} = H_{el}\{{\bf m}_i\} + H_{cl}$,
where $H_{cl}= (U/4)\sum_i {\bf m}_i^2$.
The $\{{\bf m}_i\} $ configurations follow 
the distribution
$
P\{{\bf m}_i\} \propto 
Tr_{c,c^{\dagger}} e^{-\beta (H_{el} +  H_{cl}) }. $

Within the static HS approximation $H_{eff}$ and $
P\{{\bf m}_i\} $ define a coupled fermion-local moment
problem. This is similar to the `double exchange' problem,
with the crucial difference that the moments are self generated
(and drive the Mott transition) rather than fixed in size.
Due to the fermion trace,
$P\{{\bf m}_i\}$ is not analytically calculable beyond
weak coupling.
To generate the equilibrium $\{{\bf m}_i\}$
we use Monte-Carlo sampling. Computing the energy cost of
an attempted update requires  diagonalising $H_{el}$.
To access large sizes within limited time, we
use a cluster algorithm for estimating the update cost.
We calculate the energy cost of an update by
diagonalizing a cluster (of size $N_c$, say) around
the reference site. We have extensively benchmarked this `traveling
cluster' method\cite{tca}. The static HS approach, retaining
spatial fluctuations, has found successful application
in correlated systems before \cite{dubi}.
The MC was done for lattices of size upto $N=12 \times 12 \times 12$,
with clusters of size $N_c=4 \times 4 \times 4$. We
calculate the thermally  averaged structure factor
$S({\bf q}) = \frac{1}{N^2}\sum_{ij}\langle{\bf m_{i}}
\cdot{\bf m_{j}}\rangle e^{i{\bf q}\cdot({\bf r}_i-{\bf r}_j)}$
at each temperature. The onset of rapid growth in $S({\bf q})$
at some ${\bf q} = {\bf Q}$, say, with
lowering $T$, indicates a magnetic
transition. Electronic properties (see Supplement)
are calculated by diagonalising $H_{el}$ on the full
lattice for equilibrium $\{{\bf m}_i\} $ configurations.
Since the MC ground state can be affected by annealing protocol,
wherever possible we have tested it against variational
choices of $\{{\bf m}_i\} $.

Fig.\ref{phd-fcc} shows the $U-T$ phase diagram of our model.
First focus on the magnetism at $T=0$.
($i$)~The MC based minimization,
${\delta \over {\delta {\bf m}_i}}
\langle H_{eff}\{{\bf m}_i\} \rangle = 0$,
leads to a state with $m_i=\vert {\bf m}_i \vert = 0$
for $U < U_{c1} \sim 4t$. 
($ii$)~For $U_{c1} < U < U_{c2}$,
where $U_{c2} \sim 6.7t$, the ground state involves 
finite $m_i$, with a finite width distribution 
$P(m,U) = \langle {1 \over N}
\sum_i \delta(m - \vert {\bf m}_i \vert) \rangle$, 
but with no long range spatial correlation. 
The system behaves like a spin glass with short range 
`flux like' correlations.
($iii$)~Beyond $U_{c2}$ the ground state has long range flux
like order till $U \sim 30t$, beyond which the
virtual hopping generated exchange is effectively
nearest neighbour and we 
obtain `C type' order. The C type order is indeed 
expected \cite{gvoz}
for the AF Heisenberg model on the FCC lattice.

\begin{figure}[t]
\centerline{
\includegraphics[width=8.5cm,height=6cm]{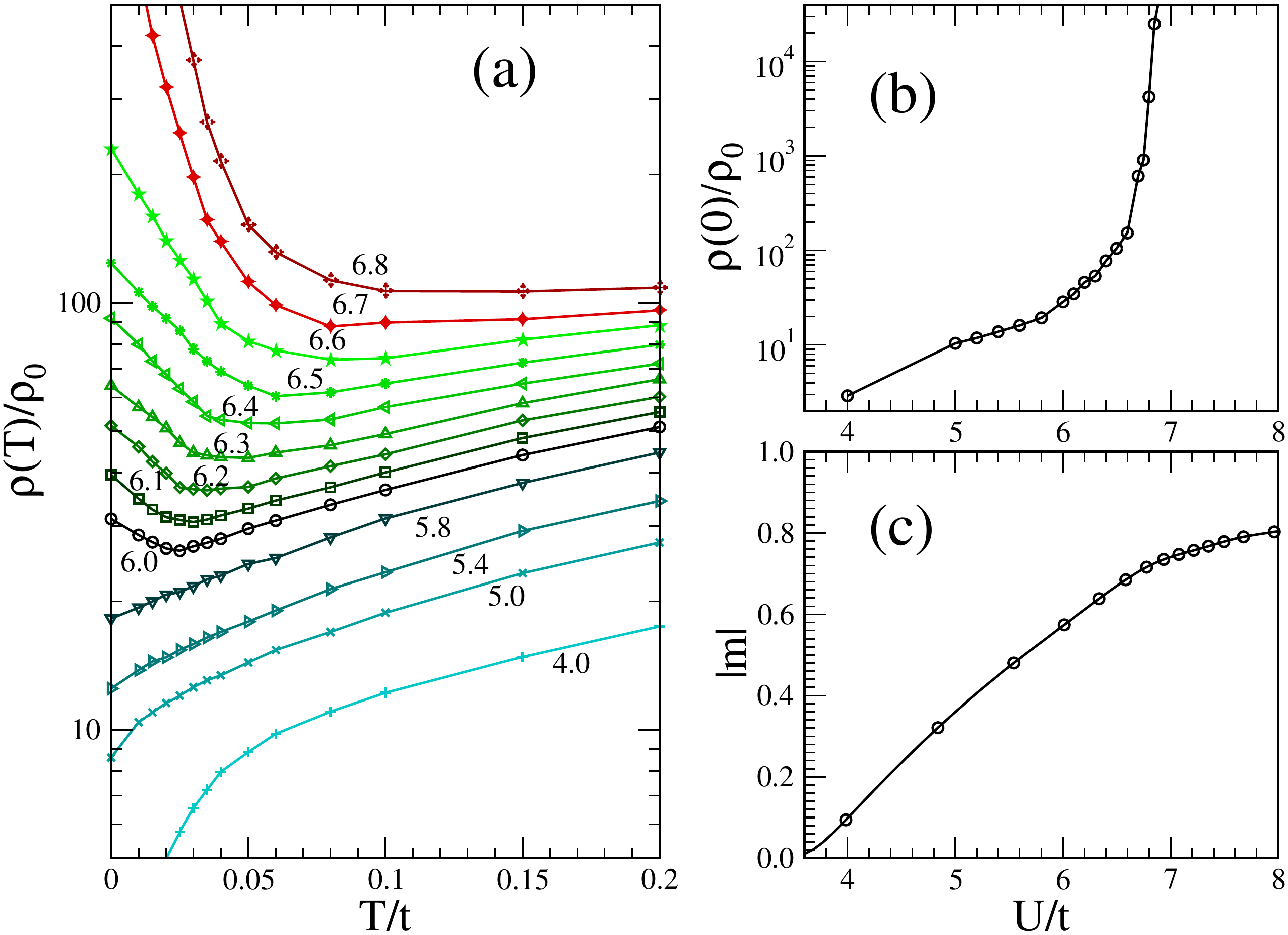}
}
\caption{\label{rho-fcc}Colour online: 
(a)~Temperature and $U$ dependence of the resistivity.
The $U/t$ values are marked in the plot.
In the PM window, $U < U_{c1}$,
the zero temperature resistivity $\rho(0)$
vanishes and
$d\rho/dT > 0$ at low $T$. In the SGM phase, $U_{c1}
< U < U_c$, $\rho(0)$ is finite, with $d\rho/dT > 0$. In
the SGI phase, {\it i.e}, $U_c < U < U_{c2}$, $\rho(0)$ is
finite, rapidly grows with $U$, and shows $d\rho/dT < 0$.
For $U > U_{c2}$ where the ground state has long range order
and a gapped spectrum, $\rho(0)$ is infinite, and $d\rho/dT < 0$.
For the weakly insulating ground states, $(U - U_c)/U_c \ll 1$,
increasing temperature leads to a crossover to $d\rho/dT > 0$
beyond a temperature we call $T_{IMT}(U)$.
(b)~The variation of the $\rho(0)$ with $U/t$.
(c)~The variation of the
average moment $m_{av}$ at $T=0$ with $U$.
Our resistivity is measured on the scale of
$\rho_0 \sim {{\hbar} a_0}/{\pi e^2}$, where $a_0$ is the lattice
spacing. For $a_0 \sim 3 \AA$ it will be $\sim 60 \mu 
\Omega$cm.
}
\end{figure}

The thermal physics deep in the AFI phase is controlled
by angular fluctuations of the local moments, ${\bf m}_i$,
about the ordered state. For $U \gg U_{c2}$ this leads to
the usual $T_c \propto t^2/U$, but with a coefficient
of $\approx 0.6$, much smaller than $\sim 1.4$
in the simple cubic case. 
At weaker interaction, for $ U_{c2} < U \lesssim 30t$, 
longer range and multi-spin couplings between the ${\bf m}_i$
become relevant and the $T_c$
deviates significantly from the $t^2/U$ asymptote.
For $U \lesssim 9t$ the size of
the local moment itself diminishes 
rapidly, due to increase in itinerancy, and the
$T_c$ falls sharply.
Below $U_{c2} \sim 6.7t$ where we
have a glassy phase we make a crude 
estimate of the `freezing temperature' from the 
MC based local relaxation 
time\cite{binder-young}, 
$\tau_{av}(T,U) = (1/N)\sum_i
\int_0^{t_{max}} dt \langle {\bf m}_i(0).{\bf m}_i(t) \rangle
$.
If the system undergoes an ordering transition, on
lowering $T$, there is a rapid growth in $\tau_{av}$
accompanied by a growth in the structure factor $S({\bf q})$
at the ${\bf q}$'s associated with long range order (LRO). For
a glass transition, one observes similar growth in
$\tau_{av}$, without any signatures in  $S({\bf q})$.
For $U > U_{c2}$ we observe LRO 
as well as a rapid increase in $\tau_{av}$ at a single
temperature 
$T_c(U)$. For the window $U_{c1} < U < U_{c2}$, however,
$\tau_{av}$ rises, at a temperature we call $T_{SG}(U)$, {\it without
associated LRO.} 
We have also `heated' the system up from $T=0$ and discovered
that any assumed ordered state is quickly destabilized while
the moments themselves survive.
$T_{SG}$ varies in the manner shown in Fig.\ref{phd-fcc}, vanishing for
$U < U_{c1}$ where there are no local moments.

\begin{figure}[b]
\centerline{
\includegraphics[width=8.0cm,height=8.0cm]{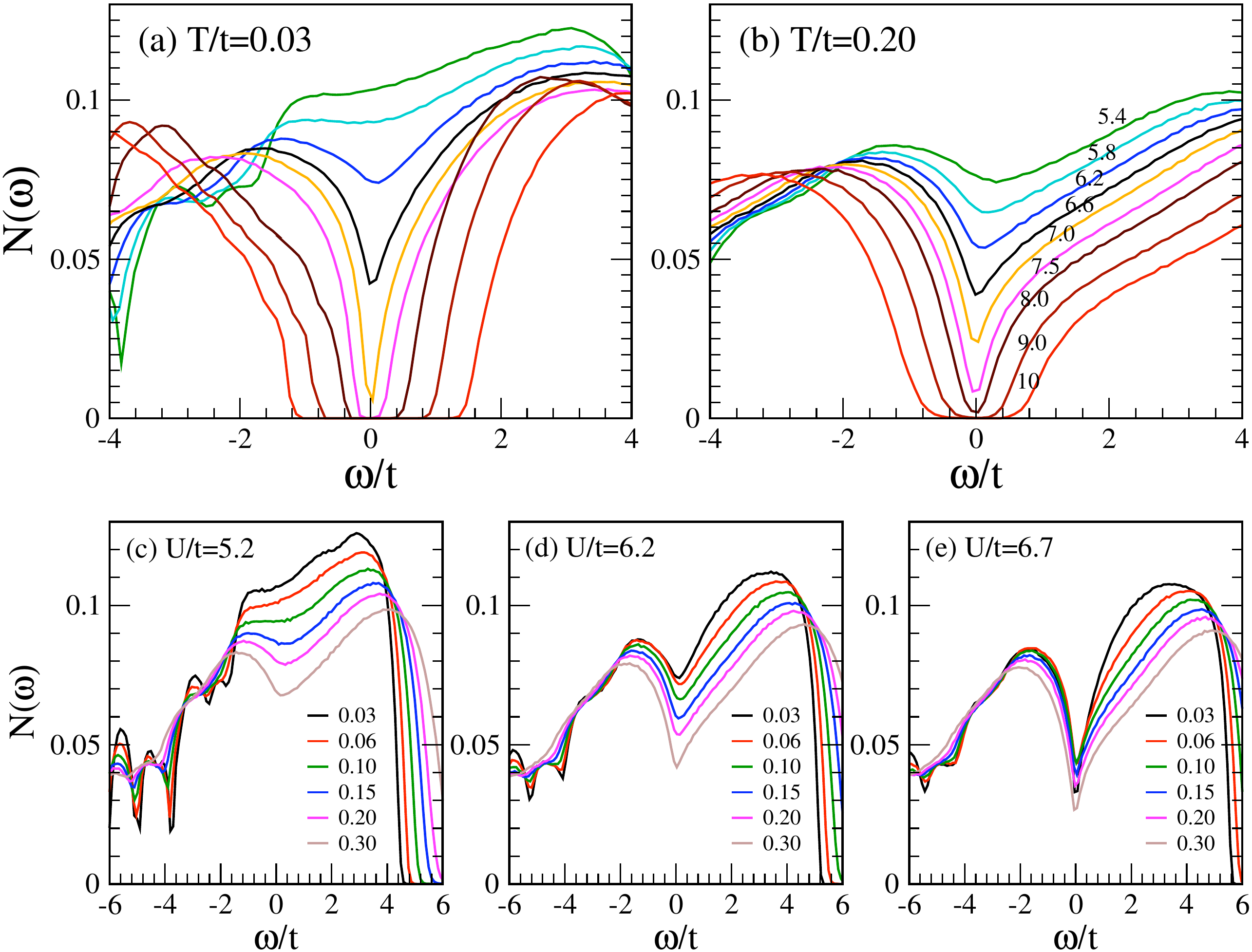}
}
\caption{\label{dos-fcc}Colour online: Density of states
$N(\omega)$.
Panel (a) shows the DOS at $T=0.03t$ for increasing $U/t$
showing the crossover from a correlated metal to an AFI
(with weak surviving order) through a wide pseudogap window.
The colour code for $U/t$ is marked in panel (b).
Panel (b) shows the DOS at $T=0.20t$ where the
crossover is between the PM and a PI through a much
wider pseudogap window.
Panels (c)-(e) show the temperature dependence at
three fixed $U$ in the spin glass window. Notice the emergence of
a thermally induced PG at weak interaction, $U=5.2t$, and
the presence of the PG at $T=0$ itself for $U=6.2t$ and $6.7t$.
Modest changes of temperature, $T \sim 0.1t$, leads to large
asymmetric shift of spectral
weight from $\omega \sim 0$ to $\omega \sim U$.}
\end{figure}
\begin{figure}[t]
\centerline{
\includegraphics[width=8.0cm,height=8.0cm]{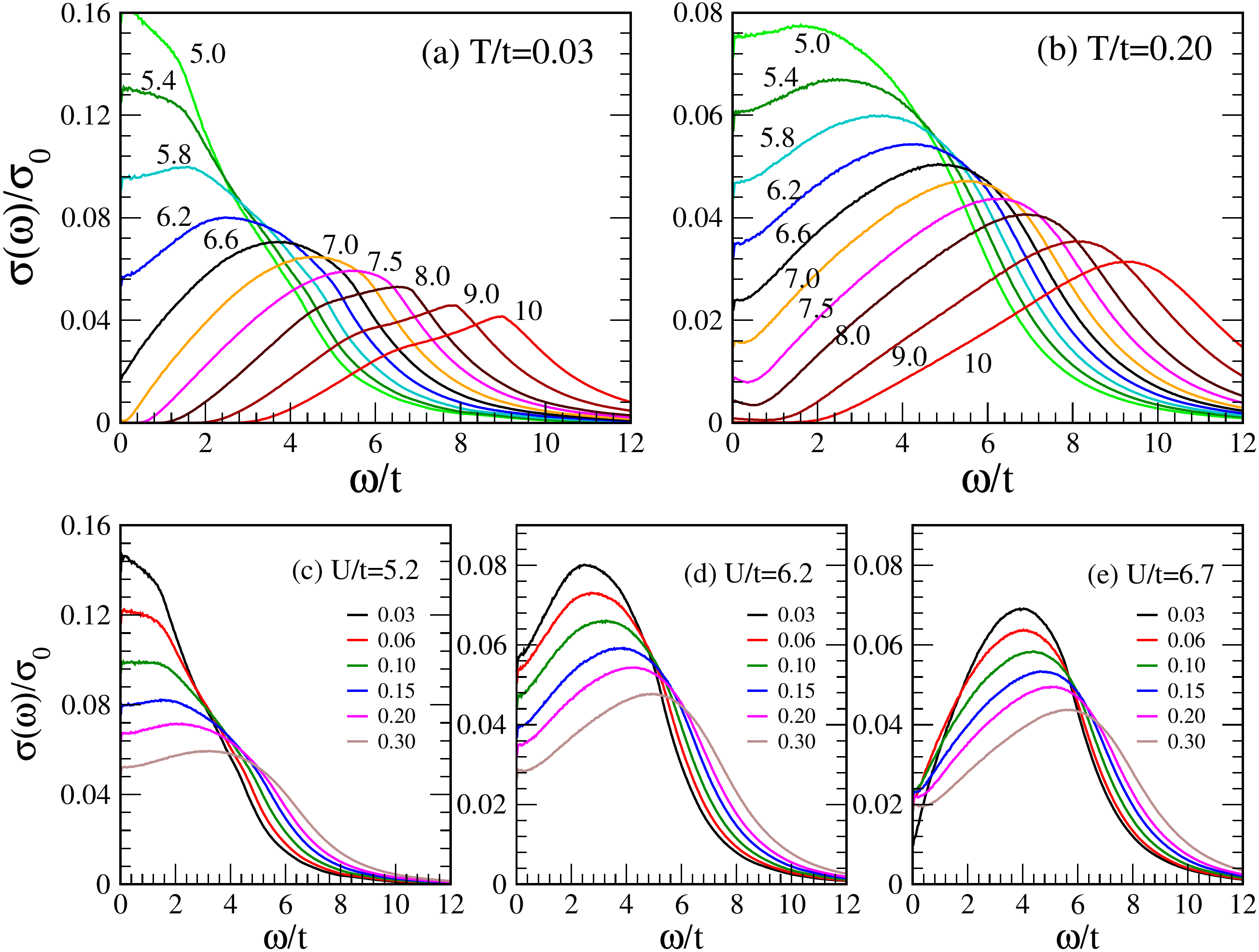}
}
\caption{\label{opt-fcc}
Colour online: Optical conductivity $\sigma(\omega)$.
Panel (a) shows $\sigma(\omega)$ at $T=0.03t$ 
for increasing $U/t$ (marked in the plot). The
response
evolves from a Drude character to the gapped spectrum of the
AFI through a non Drude regime.
Panel (b) shows similar evolution at $T=0.20t$,
from the PM to a PI.
There is no Drude peak visible down to $U=5t$, the peak locations have
moved to higher $\omega$, and the overall scale of $\sigma(\omega)$
is halved.
Panels (c)-(e) show the temperature dependence of $\sigma(\omega)$
for some $U/t$ in the spin glass window. The weak moment
system in (c) shows essentially a broadening Drude response with
increasing $T$. The larger moment system with $U/t=6.2$ in
(d) shows a non Drude response with low frequency weight suppressed
by temperature for $T > 0.03t$. Panel (e) shows a spin glass with
large $T=0$ resistivity. The very low frequency weight (on the scale
of the pseudogap) {\it increases with $T$}, while the weight 
at $\omega \sim U$ reduces with increasing $T$.
The scale $\sigma_0 = 1/\rho_0$.
}
\end{figure}

Fig.\ref{rho-fcc} shows the resistivity $\rho(T,U)$.
For $U < U_{c1}$ the resistivity $\rho(T = 0) =  0$,
and $d \rho/dT >0$. For $U > U_{c2}$ the system is gapped at $T=0$, 
$\rho(T = 0) \rightarrow \infty$ and 
$d\rho/dT <0$. These are the obvious metallic and insulating
behaviour that one expects across a correlation driven
transition.
For  $U_{c1} <  U < U_{c2} $, however, the $T=0$ resistivity is
{\it finite}, with $d\rho/dT >0$ for $U_{c1}  < U < U_c$,
where $U_c \approx 5.8t$,
and $d\rho/dT <0$ for $U_c  < U < U_{c2} $. This behaviour would
usually not be expected in a translation invariant system,
and arises because of the scattering of electrons from the
`frozen' local moments. The growth of $m_{av}(U)$,
the mean magnitude of $m_i$, 
leads to the enhanced scattering with increasing
$U$ and finally the divergence of $\rho(0)$ 
due to the opening of a gap.
The variation of $\rho(0)$ and $m_{av}$ with $U/t$ are
shown in panels (b) and (c) respectively.
We characterize the system as metallic, at a given $T$ and $U$,
when $d\rho/dT >0$, and insulating when $d\rho/dT <0$. 
With this convention, an `insulator' may have finite
spectral weight at $\omega =0$ in the optical 
conductivity $\sigma(\omega)$.

While the $U > U_{c2}$ system would have a gapped DOS, 
and the $U < U_{c1}$ case is likely to have a
featureless DOS, the glassy window in between may have unusual
spectral features. Fig.\ref{dos-fcc}.(a)-(b) shows the DOS for
varying $U/t$ at $T=0.03t$ and $T=0.20t$, respectively.
For lack of space the colour codes for $U/t$ are marked in
panel (b) only.
In panel (a), for $U \lesssim U_{c1}$ the DOS is featureless,
but for $5.4 < U/t < 7.5$ it displays a PG, and for higher
$U/t$ there is a clear gap. The large $U/t$ phase is magnetically
ordered at this temperature.
At the higher temperature in
panel (b), where there is no trace of magnetic order, 
the PG feature extends over a much larger $U/t$ window. The
weaker $U$ `metals' in (b) 
have a deeper PG compared to panel (a),
while the weak gap insulators now have a PG feature rather than
a hard gap. The evolution in panel (b) essentially illustrates 
the paramagnetic Mott transition on the FCC lattice. 

Panels \ref{dos-fcc}.(c)-(e) show the $T$ dependence of the DOS for three typical
$U/t$ in the glassy window, where the ground state has frozen
local moments. They all share the feature of thermally induced 
loss of low frequency weight which shows up at $\omega \sim U$.
There is markedly less  change with $T$ on the negative
frequency side, particularly in panels (d) and (e),
compared to positive frequencies.
This is due to the large asymmetry in the tight binding DOS 
of the FCC lattice.
There is a subtle low energy difference between panels (c)-(d)
and panel (e). In (c)-(d) the loss in low frequency weight,
within $\omega \sim \pm t$ is monotonic with $T$. In panel
(e), however, which neighbours the AFI, the low
frequency weight first {\it increases} with $T$, upto
$T \sim 0.1t$, and then again diminishes at higher temperature.

Fig.\ref{opt-fcc} shows the optics for the same parameter choice as the
DOS plots. 
Panels (a)-(b) show the evolution of $\sigma(\omega)$ across
the metal-insulator transition, between the PM and AFI at
$T=0.03t$, and between the PM and PI at $T=0.20t$. 
There is a clear window of non Drude 
response at low $T$, roughly corresponding to the
PG regime in Fig.\ref{dos-fcc}.(a). In \ref{opt-fcc}.(b) the non Drude window
in $U/t$ has increased as in Fig.\ref{dos-fcc}.(b)
with a general suppression in the magnitude of $\sigma(\omega)$.
The panels (c)-(e) show the suppression of low frequency optical
weight, with some of it appearing at $\omega \gtrsim U$. Unlike
the single particle DOS, the total 
optical weight is not conserved and varies with the kinetic
energy. At $U/t =6.7$, the very low frequency
optical response is non monotonic in $T$, showing a 
quick increase and then a gradual suppression.
This directly relates to the behaviour of $\rho(T)$ in
Fig.\ref{rho-fcc}.(a).

We have highlighted a host of magnetic, transport and spectral
features associated with Mott phenomena on the FCC lattice. 
However, like all many body methods, our approach too is
approximate and let us touch upon the possible shortcomings
before we attempt to relate our results to experiments.
Earlier papers \cite{hubb,schulz} have 
set out the formalism so we do not enter into it again here.

The ground state that we access through MC
is equivalent to the UHF result, but {\it with no
assumptions about translational symmetry}.
It is easy to see some of the qualitative effects  
of  dynamical fluctuations  
in  ${\bf m}_i$ and $\phi_i$, that
we have neglected,  at $T=0$.
These  would (i)~convert the $U < U_{c1}$ 
PM to a correlated metal, (ii)~introduce quantum spin
fluctuations in the large $U$ AFI, and (iii)~possibly
shift $U_{c1}$ to a larger value (since the correlated
metal competes better with the local moment phase). 
The intermediate window `spin glass' that appears within the
static approximation might be converted to a spin liquid
with slowly fluctuating moments. 
A recent calculation on the triangular lattice
demonstrates how longer range and multi-spin interactions
arise on a frustrated Mott insulator and can lead to
a spin liquid ground state \cite{pcut}.

Our approach captures the correct thermal 
fluctuations of the ${\bf m}_i$, without any assumption about
LRO in the background. This in
turn allows us to capture a $T_c$ that has the qualitatively
correct $U$ dependence. With growing temperature, but staying at
$T \ll U$, these classical thermal fluctuations should reasonably
describe the magnetic background, and its effect on the electrons.

While our solution of the FCC Mott problem involves
approximation, and real materials usually require 
interactions and degrees of freedom beyond the 
Hubbard description, our results suggest the
organization of a broad class of experiments.
(i)~We find that a non coplanar spin configuration
dominates the Mott phase, and pressure induced metallisation
leads to a weak moment `spin frozen' state with 
short range non coplanar correlations.
This is consistent 
\cite{148-Nb-flux,c60-nmr,c60-phd} 
with observations on GaTa$_4$Se$_8$,  FCC
A$_3$C$_{60}$, and 
double perovskite materials. 
(ii)~Beyond the pressure driven 
IMT the materials exhibit
\cite{148-res-elm,pyr-Mo,pyr-Ir-jpsj,pyr-Ir-cdyn}
very high $\rho(0)$, and $d\rho/dT < 0$.
Our results
show how this can arise from the presence of 
disordered local moments strongly
coupled to the itinerant electrons, leading to large
scattering. In the Ga cluster materials these
local moments emerge
from correlation effects, while in the pyrochlores\cite{pyr-Mo}
they are already present as localised $a_{1g}$ electrons.
(iii)~The very recent optical measurement 
\cite{148-opt-phu} in GaTa$_4$Se$_8$
shows pronounced non Drude
character in the metal near the IMT. This is consistent
with our optics results for the PM to PI transition.
In fact we have a {\it quantitative} description
of the Mott transition in these 
materials \cite{cluster-unpub} 
within our present framework.


The correspondence above allows us to make two
concrete predictions: (a)~The frustrated Mott systems
should have a 
wide pseudogap regime beyond the insulator-metal
transition, 
persisting to $T=0$. These should be visible in
tunneling and photo-emission spectra. (b).~The 
thermally induced shift of single particle spectral
weight would be {\it extremely asymmetric}. Weight at
low positive frequencies is shifted to the scale of
$\omega \sim U$, while the negative frequency 
spectrum remains almost unaffected.

We have not probed the anomalous Hall response due to
flux like correlations and plan to present them separately.
Some of the FCC Mott materials have a superconducting
instability at very low temperature, $\sim 10$K. We have
not treated that aspect, and need a method that retains the
dynamics of the ${\bf m}_i$ to do so.

{\it Conclusions:}
We have provided the first comprehensive study of the Mott
transition on the geometrically frustrated FCC lattice. 
The magnetic frustration leads to a 
seemingly `two fluid' state of itinerant electrons and 
disordered local moments 
between the paramagnetic metal and the 
antiferromagnetic insulator.
The disordered phase involves a large residual
resistivity, non Drude optical response, and a single
particle pseudogap. 
The temperature and interaction dependence that we
uncover allows a common conceptual scheme for a wide
variety of materials.

We acknowledge use of the HPC clusters at HRI. PM acknowledges
support from a DAE-SRC Outstanding Research Investigator grant.

\bibliographystyle{unsrt}

\newpage
\section*{Supplementary information:}

\subsection{Magnetic phases on FCC lattice}

\begin{figure}[ht]
\includegraphics[width=7.0cm,height=7cm]{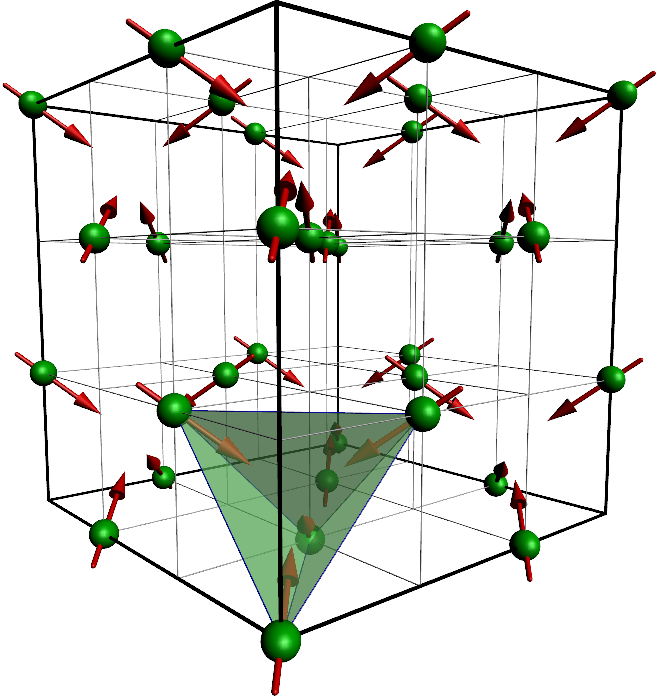}\\
\includegraphics[width=7.0cm,height=7cm]{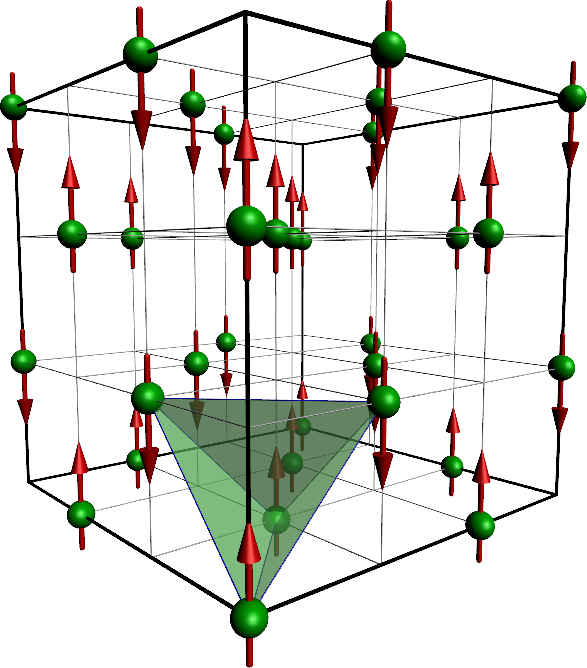}
\caption{\label{conf}Color online:
  The `flux' (top) and `C-type' (bottom) phases, drawn on cubic lattice motif.
  The green spheres are the FCC lattice points, and arrows indicate the direction
  of ${\bf m_{i}}$ moments. The tetrahedra highlights the geometrical frustration.
}
\end{figure}

In the cubic lattice notation, the primitive lattice translation vectors for FCC are
${\bf A_1}$ $=a(0,1,1)$,${\bf A_2}$ $=a(1,0,1)$,${\bf A_3}$ $=a(1,1,0)$, so that the
nearest neighbour distance is $d_{fcc}=\sqrt{2}a$. All the points on FCC lattice
are expressed in integer units of these,
i.e., $X = \sum_{i=1}^{3}n_iA_i$ = $a(n_2+n_3,n_3+n_1,n_1+n_2)$.
Each site has 12 neighbours $X+\delta$, where $\delta$=$(\pm 1,\pm 1,0),(\pm 1,0,\pm 1),(0,\pm 1,\pm 1)$.
The `flux' phase mentioned in the text, is described by the following formula
$$
{\bf m}(X) = m(e^{iQ_1\cdot X},e^{iQ_2\cdot X},e^{iQ_3\cdot X})/\sqrt{3}
$$
Where, $m$ is magnitude of the vector, and $Q_1=(\frac{\pi}{a},0,0)$, $Q_2=(0,\frac{\pi}{a},0)$, $Q_3=(0,0,\frac{\pi}{a})$.
Its a non-coplanner phase. On the other hand, the `C-type' phase is a collinear one,
described by the formula
$$
{\bf m}(X) = m(0,0,e^{iQ_c\cdot X})
$$
Where, $Q_c=(\frac{\pi}{a},\frac{\pi}{a},0)$. It consists of alternating ferromagnetic lines. Both the `flux'
and `C-type' phases are shown in Fig.\ref{conf} in the top and bottom panel respectively.

\subsection{Derivation of the effective Hamiltonian}
Consider the single band Hubbard model as starting point
\begin{eqnarray}
H &=& \sum_{\langle ij\rangle\sigma}t_{ij} 
c^{\dagger}_{i\sigma}c_{j\sigma} -\mu\sum_i n_i
  + U\sum_{i}n_{i\uparrow}n_{i\downarrow} \cr
&=& H_0 -\mu\sum_i n_i 
+ U\sum_{i}n_{i\uparrow}n_{i\downarrow} \nonumber
\end{eqnarray}
and implement a rotation invariant decoupling of
the Hubbard term as follows. First, we write
$$
n_{i\uparrow}n_{i\downarrow}=
\frac{n_i^2}{4}-(\vec{s}_i\cdot {\hat {\bf m}_i})^2
$$
where $n_i =n_{i\uparrow}+n_{i\downarrow}$ is the charge 
density, 
$\vec{s}_{i}=\frac{1}{2}\sum_{\alpha,\beta}
c^{\dagger}_{i\alpha}\vec{\sigma}_{\alpha\beta}c_{i\beta}=2\vec{\sigma}_i$
is the local electron spin operator, and ${\hat {\bf  m}_i}$ is an
{\it arbitrary unit vector}.

The partition function of the Hubbard model is
\begin{eqnarray}
Z & = & \int D[c,\bar{c}]e^{-S} \cr
S & =&  \int_{0}^{\beta}d \tau {\cal L}(\tau) \cr
{\cal L} & = & 
\sum_{i\sigma}
\bar{c}_{i\sigma}(\tau)\partial_{\tau}c_{i\sigma}(\tau)
+ H(\tau) 
\nonumber
\end{eqnarray}
We introduce two space-time varying auxiliary fields
for a Hubbard-Stratonovich transformation:
(i)~$\phi_{i}(\tau)$ coupling to charge density, and 
(ii)~$\Delta_{i}(\tau) {\hat {\bf m}_i}(\tau)={\bf m}_i(\tau)$
coupling to electron spin density ($\Delta_i$ is real positive).
This allows us to define an SU(2) invariant HS transformation\cite{weng,dupuis},
$$
  e^{U n_{i\uparrow}n_{i\downarrow}}  = 
  \int \frac{d\phi_{i} d {\bf m}_i}{4\pi^{2}U}
  e^{ 
\left( 
\frac{\phi_{i}^2}{U}+i\phi_{i} n_i
+ \frac{\bf {m}_i^2}{U}-2{\bf m}_i\cdot\vec{s}_{i}
\right)
}
$$
The partition function now becomes:
\begin{eqnarray}
  Z &=& 
\int\prod_{i}\frac{d\bar{c_{i}}dc_{i}d\phi_{i} d{\bf m}_{i}}{4\pi^{2}U}
  e^{\left(-\int_{0}^{\beta}{\cal L}(\tau)\right)} \cr
{\cal L}(\tau) &=& 
\sum_{i\sigma}
\bar{c}_{i\sigma}(\tau)\partial_{\tau}c_{i\sigma}(\tau)
+ H_0(\tau)
+ {\cal L}_{int}(\phi_i(\tau), {\bf m}_i(\tau)) \cr
{\cal L}_{int} &=& \sum_{i}\left[\frac{\phi_{i}^2}{U}+
i\phi_{i} n_i+
 {{\bf m}_i^2 \over U}  - 2 {\bf m}_{i}\cdot\vec{s}_{i}\right]
\nonumber
\end{eqnarray}

As mentioned in the text, to make progress we use two 
approximations: (i)~neglect the time ($\tau$) dependence
of the HS fields, (ii)~replace the field $\phi_i$ by its
saddle point value $(U/2)\langle n_i \rangle =U/2$,
since the important low energy fluctuations arise
from the ${\bf m}_i$. 
Simplifying the action with above substitution, we get
the effective Hamiltonian 
$$
H_{eff} = H_{0} - {\tilde \mu} \sum_i n_i 
- \sum_{i} {\bf m}_i\cdot\vec{\sigma}_{i}
+ \sum_{i} {{\bf m}_i^2 \over U} 
$$
where ${\tilde \mu} = \mu - U/2$.
For the sake of convenience
we rescale ${\bf m}_i \rightarrow \frac{U}{2} {\bf m}_i $, so that 
the ${\bf m}_i$ is dimensionless.
This leads to the effective Hamiltonian that we used in the text:
$$
H_{eff} = H_{0} - {\tilde \mu} \sum_i n_i 
- {U \over 2} \sum_{i} {\bf m}_i\cdot\vec{\sigma}_{i}
+ {U \over 4} \sum_{i} {\bf m}_i^2  
$$
The partition function can be written in terms of $H_{eff}$
$$
Z = \int {\cal D} {\bf m}_i Tr_{c,c^{\dagger}} e^{-\beta H_{eff}}
$$
For a given configuration $\{ {\bf m}_i\}$ 
the problem is quadratic in the fermions,
while the configurations themselves are obtained by a MC sampling as discussed in the text.

\subsection{Optical conductivity}
The conductivity of is calculated as follows
(ref.\cite{allen}), using the Kubo formula:
\begin{eqnarray}
  \sigma^{xx}(\omega) &=& \frac{A}{N}\sum_{\alpha,\beta}
  { {n_{\alpha}-n_{\beta}} \over {\epsilon_{\beta}-\epsilon_{\alpha}} } 
  |\langle \alpha|J_x|\beta\rangle|^2
  \delta(\omega-(\epsilon_{\beta}-\epsilon_{\alpha}))
  \nonumber\\
  J_x&=& -it\sum_{i,\sigma}\left[(c^{\dagger}_{i,\sigma}c_{i+\hat{x}+\hat{y},\sigma}-\textrm{hc})\right]\nonumber
\end{eqnarray}

Where, $J_x$ is the current operator and, the coefficient $A=\frac{\sigma_0}{2a}=\frac{\sigma_0}{\sqrt{2}d_{fcc}}$.
$\sigma_{0}$=$\frac{\pi e^2}{\hbar}$ is the scale for conductivity with dimension of conductance.

$n_{\alpha}=f(\epsilon_{\alpha})$ is the Fermi function,
and $\epsilon_{\alpha}$ and $|\alpha\rangle$ are 
respectively the single particle eigenvalues and eigenstates of
$H_{eff}$ in a given background \{${\bf m}_i$\}. $N$ is the number of sites ($N=12^3$ in our results).
The results we show in the text are 
averaged over equilibrium MC configurations.

The d.c conductivity is the $\omega \rightarrow 0$ limit of 
the result above. The experimental value of $d_{fcc}$ for the GaTa$_4$Se$8$ cluster compound
is $\sim 4.3\AA$.

\bibliographystyle{unsrt}

\begin{thebibliography}{99}
\bibitem{chem-press}
  The effective `pressure' can also be of chemical origin,
  driven by change in ionic radius.
\end{thebibliography}

\end{document}